\pdfoutput=1
\documentclass[fleqn,12pt]{wlscirep}
\usepackage[utf8]{inputenc}
\usepackage[T1]{fontenc}
\usepackage{setspace}
\usepackage[mathlines]{lineno}
\usepackage{threeparttable}

\makeatletter
\renewcommand{\@maketitle}{%
{%
\thispagestyle{empty}%
\vskip-36pt%
{\raggedright\sffamily\bfseries\fontsize{20}{25}\selectfont \@title\par}%
\vskip10pt
{\raggedright\sffamily\fontsize{12}{16}\selectfont  \@author\par}
\vskip18pt%
}%
}
\makeatother

\title{Patched-Wall Quasistatic Cavity Resonators for 3-D Wireless Power Transfer}

\author[1,*]{Takuya~Sasatani}
\author[1]{Yoshihiro Kawahara}
\affil[1]{Department of Electrical Engineering and Information Systems, The University of Tokyo, Tokyo, Japan}

\affil[*]{sasatani@g.ecc.u-tokyo.ac.jp}

\begin{document}
\singlespacing
\flushbottom
\maketitle
\thispagestyle{empty}

\section*{Abstract}
Traditional wireless power transfer (WPT) systems are largely limited to 1-D charging pads or 2-D charging surfaces and therefore do not support a truly ubiquitous device-powering experience. Although room-scale WPT based on multimode quasistatic cavity resonance (QSCR) has demonstrated full-volume coverage by leveraging multiple resonant modes, existing high-coverage implementations require obstructive internal conductive structures, such as a central pole. Here, we present a new structure, termed the patched-wall QSCR, that eliminates such internal obstructions while preserving full-volume coverage. By using conductive wall segments interconnected by capacitors, the proposed structure supports two complementary resonant modes that cover both the peripheral and central regions without obstructions within the charging volume. Electromagnetic simulations show that, by selectively exciting these two resonant modes, the proposed structure achieves a minimum power-transfer efficiency of 48.1\% throughout the evaluated interior volume of a 54~m$^3$ enclosure while preserving an unobstructed charging space.

\section*{Introduction}
Electronic devices, including cell phones, sensor nodes, and robots, are now pervasive in industrial and living spaces~\cite{weiser_computer_1991, xu_internet_2014}. However, these devices remain powered primarily through wired connections or batteries, which impose recurring charging and replacement burdens and, for disposable batteries, an environmental cost. Wireless power transfer (WPT) is increasingly used for mobile-device charging, yet achieving wide spatial coverage, high delivered power, and high efficiency simultaneously remains challenging. Microwave approaches can provide long transfer distances, but they typically suffer from limited delivered power and low efficiency~\cite{garnica_wireless_2013}. Conventional cavity-resonator-based approaches can provide three-dimensional coverage, but they rely on exposed electric fields within the resonant volume, resulting in strong interactions with dielectric objects~\cite{chabalko_resonant_2014,tamura_cavity-resonance-enabled_2026}. Magnetoquasistatic approaches enable high-power, high-efficiency power transfer, but they are largely restricted to 1-D charging cradles or 2-D charging surfaces because of their limited range~\cite{Kurs2009, Hui2014, Shinohara2011, Sample2011, wu_wireless_2022, Guo2017, younesiraad_analysis_2018, paul_fast_2018, ahmed_optimized_2025}.

Recent work on quasistatic cavity resonance (QSCR) has demonstrated a route toward 3-D, wide-range WPT in the magnetoquasistatic regime by leveraging surface currents on a conductive enclosure~\cite{Chabalko2017, sasatani_geometry-based_2022,10897905}. These systems use conductive sheets, lumped capacitors, and additional conductive structures to allow magnetic fields to permeate the cavity interior while confining electric fields to the lumped capacitors, thereby enabling high-efficiency, high-power transfer over large volumes. Furthermore, subsequent work on multimode QSCR showed that multiple resonant modes can be used to provide full-volume coverage~\cite{sasatani_multimode_2017,sasataniRoomscaleMagnetoquasistaticWireless2021,rong_cavity_2025}. However, existing high-coverage implementations require conductive structures within the room interior, which obstruct normal use of the space.

Here, we propose the patched-wall QSCR, a resonator structure that achieves full-volume coverage by preserving the pole-independent (PI) mode while replacing the pole-dependent (PD) mode with a new wall-based resonance, termed the surface-Helmholtz (SH) mode (Fig.~\ref{fig_overview}). Prior multimode QSCR designs relied on a PI mode to cover regions near the walls and a PD mode to cover the central region of the volume. Because the PI mode is largely unaffected by the central pole, it is preserved in the proposed structure. In contrast, the PD mode requires a physical central pole to guide current and shape the magnetic field in the center of the volume. To replace this mode, we introduce the SH mode, synthesized using segmented cavity walls interconnected by capacitors to shape current flow along the ceiling and floor. This mode generates a central magnetic field using the floor and ceiling as a pair of current loops, thereby eliminating the need for internal conductive structures. As a result, wide-area high-efficiency power delivery can be achieved while maintaining a completely unobstructed charging space. Using finite-element simulations, we evaluate the performance of the proposed system, modeled as a 54~m$^3$ enclosure, throughout its interior volume and show that selective excitation of the two modes substantially suppresses low-efficiency regions, thereby enabling unobstructed wide-area 3-D wireless power coverage.

\section*{Results}
\subsection*{Resonator design and mode structure}

\begin{figure}[t]
\centering
\includegraphics[width=\linewidth]{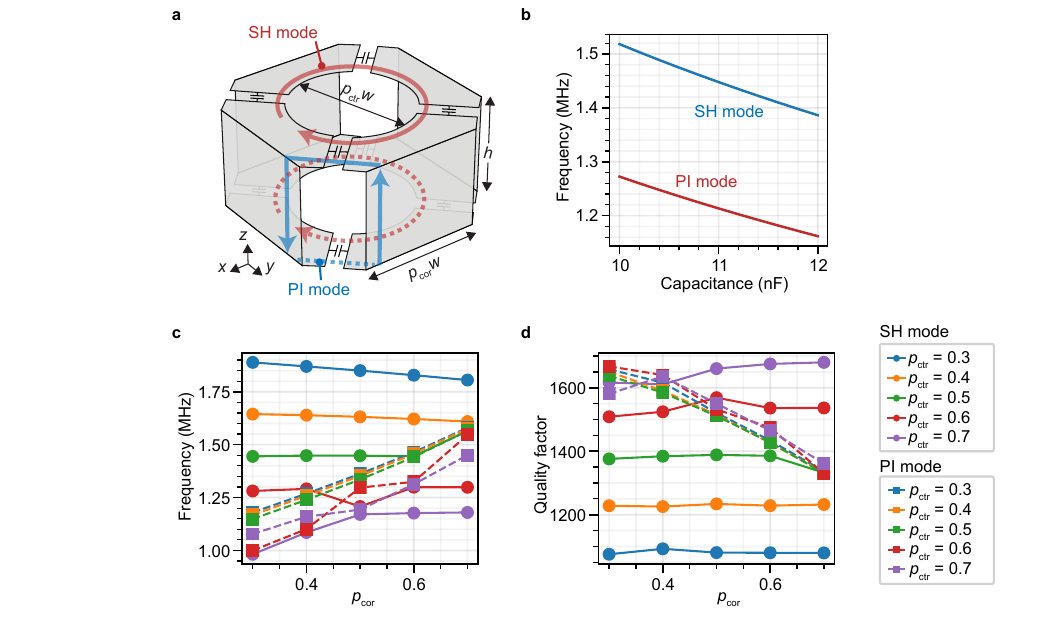}
\caption{Geometry and tunability of the proposed patched-wall resonator. (a) Resonator structure, dimensions, and qualitative current-loop paths associated with the SH and PI modes. (b) Resonant frequency versus capacitance for the two supported modes. (c) Resonant frequency versus geometric parameters $p_{\mathrm{ctr}}$ and $p_{\mathrm{cor}}$. (d) Simulated quality factor versus $p_{\mathrm{ctr}}$ and $p_{\mathrm{cor}}$ for the SH and PI modes. These results show that the operating frequencies can be tuned through both lumped capacitance and structural geometry while maintaining high quality factors.}
\label{fig_structure}
\end{figure}

The proposed patched-wall resonator consists of exterior surfaces partitioned into conductive sheet ``patches'' interconnected by capacitors and strategically placed apertures~(Fig.~\ref{fig_structure}a). Specifically, apertures are located at each corner of the room and at the centers of the ceiling and floor. The remaining metal enclosure is divided into four units coupled through lumped-element capacitors. In this work, the concept is implemented in simulation as a $4.9 \times 4.9 \times 2.3$~m$^3$ enclosure. In contrast to single-mode QSCR and prior multimode QSCR designs, in which a continuous conductive wall serves as the return path for current flowing through a central pole, the proposed architecture supports current paths that loop within the walls, floor, and ceiling. The capacitors serve two primary functions: they tune the structure to a resonant frequency in the magnetoquasistatic regime and determine the distribution of surface currents and the resulting magnetic field.

\begin{figure}[t]
\centering
\includegraphics[width=\linewidth]{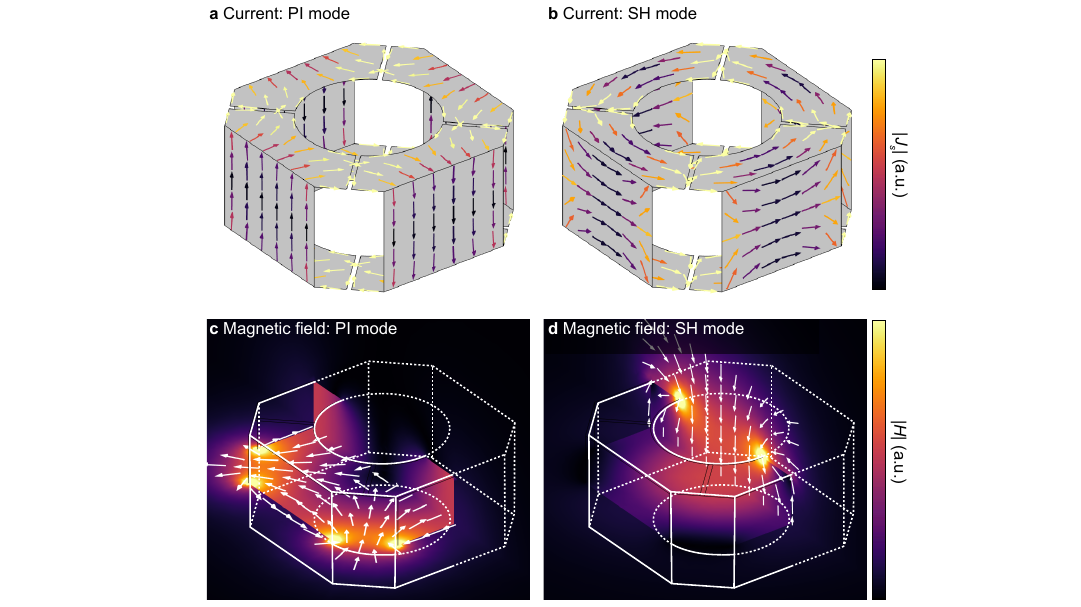}
\caption{Two complementary resonant modes of the patched-wall quasistatic cavity resonator. Simulated surface-current distributions and magnetic-field patterns are shown for the two supported modes: (a) current distribution of the pole-independent (PI) mode, (b) current distribution of the surface-Helmholtz (SH) mode, (c) magnetic-field magnitude for the PI mode, and (d) magnetic-field magnitude for the SH mode. The PI mode concentrates magnetic flux near the periphery of the cavity, whereas the SH mode produces a strong central field without requiring an internal conductive pole.}
\label{fig_overview}
\end{figure}

The system supports two complementary resonant modes that together provide coverage throughout the interior volume. Prior work achieved full-volume coverage using a pole-independent (PI) mode for peripheral regions and a pole-dependent (PD) mode for the center~\cite{sasatani_multimode_2017,sasataniRoomscaleMagnetoquasistaticWireless2021}. The proposed patched-wall QSCR follows the same principle while eliminating the need for a physical conductive pole. The PI mode remains consistent with prior multimode QSCR implementations, whereas the surface-Helmholtz (SH) mode primarily uses surface currents on the ceiling and floor to synthesize a field distribution analogous to that of a Helmholtz coil~(Fig.~\ref{fig_overview})~\cite{10557310}. This configuration consists of vertically separated current loops that produce a relatively uniform magnetic field in the central region.

\subsubsection*{Pole-independent (PI) mode}
The PI mode is excited by inducing currents that flow primarily along the vertical wall segments. As demonstrated in previous work, this mode produces magnetic-field patterns concentrated near the room boundaries and corners, thereby providing coverage in regions where center-focused modes typically exhibit nulls~\cite{sasatani_multimode_2017,sasataniRoomscaleMagnetoquasistaticWireless2021}. In the patched-wall architecture, this mode is realized by exciting the vertical gaps between adjacent wall patches. Because this mode does not rely on a central current path, its field distribution is largely unaffected by the additional apertures at the centers of the ceiling and floor.

\subsubsection*{Surface-Helmholtz (SH) mode}
The SH mode is designed to provide a strong magnetic field at the center of the enclosure, serving as a pole-free replacement for the traditional PD mode. In prior multimode QSCR designs, the PD mode relies on current flowing vertically through a central pole to generate a radial magnetic field. In the patched-wall configuration, the SH mode is excited by inducing current loops that circulate along the floor and ceiling in the same direction. This current distribution acts as a large-scale, surface-integrated Helmholtz pair. As a result, it effectively mitigates $z$-axis dependence and enables a highly uniform magnetic-field distribution throughout the central volume.

\subsubsection*{Resonant mode excitation and frequency tuning}

The combination of these two modes distributes magnetic flux throughout the charging volume. By switching the excitation position and frequency, the system can toggle between the SH and PI modes to accommodate the receiver position.

For deployment, it is desirable to tune the resonant frequency of each mode to a specific target without altering the interior space, which is typically constrained by the application. To this end, we demonstrate that (a) the resonant frequencies shift predictably with the lumped capacitance values, and (b) the relative difference between these frequencies can be adjusted through geometric parameters. Here, $p_{\mathrm{ctr}}$ denotes the relative size of the center apertures on the floor and ceiling, and $p_{\mathrm{cor}}$ denotes the relative size of the corner apertures, both normalized to the room width. Variations in these geometric parameters leave the field distributions largely unchanged, and their influence on volumetric wireless power coverage is addressed in subsequent sections through efficiency-coverage analysis.

To examine these effects and assess tunability, we performed finite-element eigenvalue simulations of the enclosure (see Methods). Fig.~\ref{fig_structure}b shows the resonant frequencies of both modes as a function of capacitance, with the geometric parameters fixed at $p_{\mathrm{cor}} = p_{\mathrm{ctr}} = 0.6$. Figs.~\ref{fig_structure}c and~\ref{fig_structure}d further show the resonant frequencies and quality factors ($Q$) over a sweep of geometric parameters. The results indicate that the resonant frequencies can be adjusted to some extent through structural geometry while maintaining high quality factors. Together with the inherent tunability of the capacitors, this demonstrates the system's ability to reach the desired target frequencies. Furthermore, although uniform capacitance across the eight specified gaps is assumed here for simplicity, adding capacitors to specific regions to independently tune a single resonant mode remains a viable approach for more flexible frequency tuning.

\begin{figure}[t]
\centering
\includegraphics[width=\linewidth]{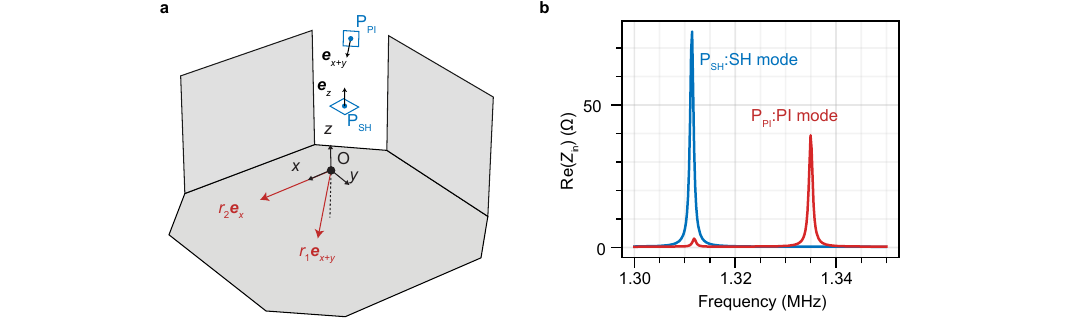}
\caption{Selective mode excitation in the proposed resonator. (a) Positions of the drive coils used for mode excitation ($\mathrm{P}_\mathrm{PI}$ and $\mathrm{P}_\mathrm{SH}$) and the coordinate system used for the receiver-position-based efficiency evaluation. (b) Simulated real part of the input impedance as a function of frequency for the two drive coil locations. The results demonstrate that varying the drive coil position and operating frequency allows for the selective excitation of the PI and SH modes.}
\label{fig_excitation}
\end{figure}

Finally, Fig.~\ref{fig_excitation} shows the input impedance for $p_{\mathrm{cor}} = p_{\mathrm{ctr}} = 0.6$ and $C=11$ nF, with the drive coils placed at two distinct positions, $\mathrm{P}_\mathrm{PI}$ and $\mathrm{P}_\mathrm{SH}$. The results demonstrate that the two modes can be selectively excited by varying the excitation location and operating frequency.

\subsection*{Power transfer efficiency}

Following the design principles of the patched-wall QSCR, we evaluate the resulting wireless power transfer efficiency throughout the 3-D interior using coupled-mode theory (CMT), and validate the CMT predictions against full-wave S-parameter simulations at representative receiver positions (see Methods).

\begin{figure}[t]
\centering
\includegraphics[width=\linewidth]{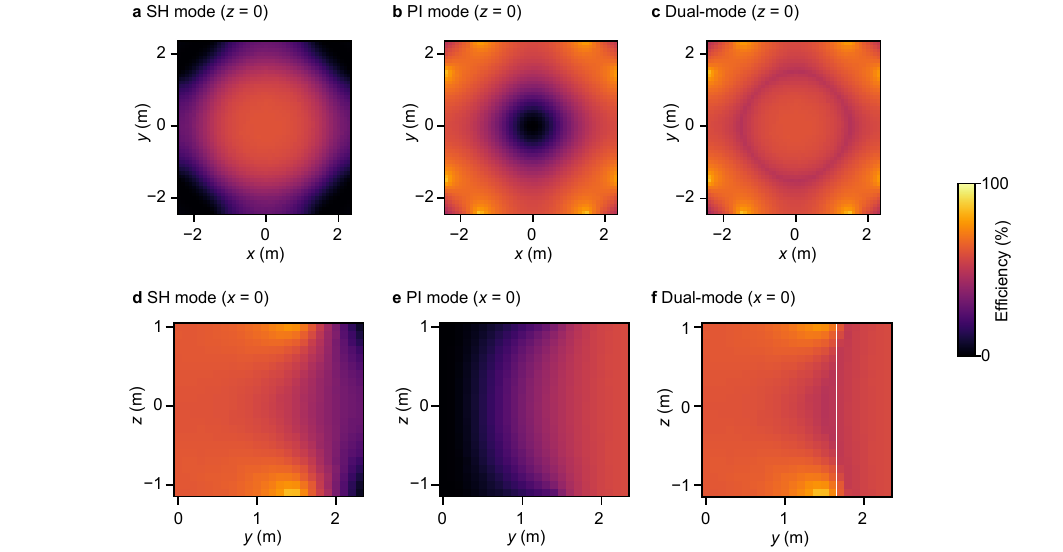}
\caption{Volumetric wireless power transfer efficiency of the patched-wall resonator. Simulated efficiency distributions are shown for the representative case of $p_{\mathrm{cor}} = p_{\mathrm{ctr}} = 0.6$: (a)--(c) efficiency on the $z=0$ plane for SH mode, PI mode, and dual-mode operation, respectively; (d)--(f) efficiency on the $x=0$ plane for SH mode, PI mode, and dual-mode operation, respectively. Dual-mode excitation combines the complementary coverage of the two resonant modes and suppresses low-efficiency regions throughout the charging volume.}
\label{fig_efficiency_pos}
\end{figure}

Fig.~\ref{fig_efficiency_pos}a--f shows the SH, PI, and dual-mode operation on the $z=0$ and $x=0$ planes for the representative case of $p_{\mathrm{cor}} = p_{\mathrm{ctr}} = 0.6$. The SH mode provides strong power transfer in the central region with relatively weak $z$-dependence, whereas the PI mode complements it by maintaining higher efficiency near the room boundaries. When the two modes are combined by selecting the more efficient mode at each receiver position (see Methods), the low-efficiency regions of each individual mode are substantially suppressed, resulting in a much more uniform coverage distribution across the charging volume.

\begin{figure}[t]
\centering
\includegraphics[width=\linewidth]{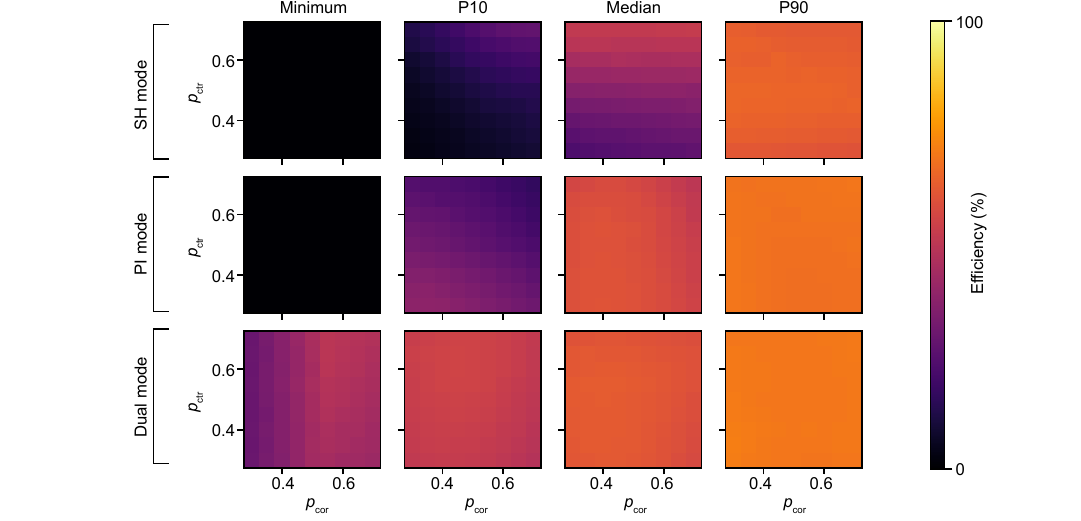}
\caption{Coverage statistics of the simulated power-transfer efficiency as the geometric parameters are varied. The rows correspond to SH-mode, PI-mode, and dual-mode operation, while the columns show the minimum, 10th percentile (P10), median, and 90th percentile (P90) efficiencies within the evaluated volume. The maps summarize how the geometric parameters $p_{\mathrm{cor}}$ and $p_{\mathrm{ctr}}$ influence volumetric coverage.}
\label{fig_efficiency_stats}
\end{figure}

To further evaluate robustness to geometry, Fig.~\ref{fig_efficiency_stats} summarizes the minimum, 10th percentile (P10), median, and 90th percentile (P90) efficiencies within the evaluated volume as the geometric parameters $p_{\mathrm{cor}}$ and $p_{\mathrm{ctr}}$ are varied. Across the parameter sweep, SH-only and PI-only operations exhibit minima close to zero, indicating that each mode alone retains weak-coverage regions despite geometric tuning. In contrast, dual-mode operation substantially improves the lower-tail coverage statistics, raising both the minimum and P10 efficiencies while maintaining high median and P90 values over a broad range of geometries. Table~\ref{tab:mode_statistics} provides a numerical summary for the representative case. For $p_{\mathrm{cor}} = p_{\mathrm{ctr}} = 0.6$, dual-mode operation increases the minimum efficiency to 48.1\% and the P10 efficiency to 53.9\% while maintaining high median and P90 values. These results show that the complementary SH and PI field distributions, rather than geometric tuning alone, are essential for achieving wide-area unobstructed coverage.

\begin{table}[t]
\centering
\caption{Volumetric efficiency statistics for each operating mode.}
\label{tab:mode_statistics}
\renewcommand{\arraystretch}{1.35}
\setlength{\tabcolsep}{10pt}
\begin{tabular}{lccccc}
\toprule
Mode & Min & P10 & Median & P90 & Max \\
\midrule
SH   & 0.0\%  & 18.8\% & 45.7\% & 63.8\% & 77.4\% \\
PI   & 0.0\%  & 22.4\% & 56.4\% & 68.9\% & 78.6\% \\
Dual & 48.1\% & 53.9\% & 61.4\% & 69.8\% & 78.6\% \\
\bottomrule
\end{tabular}
\end{table}

\begin{figure}[t]
\centering
\includegraphics[width=\linewidth]{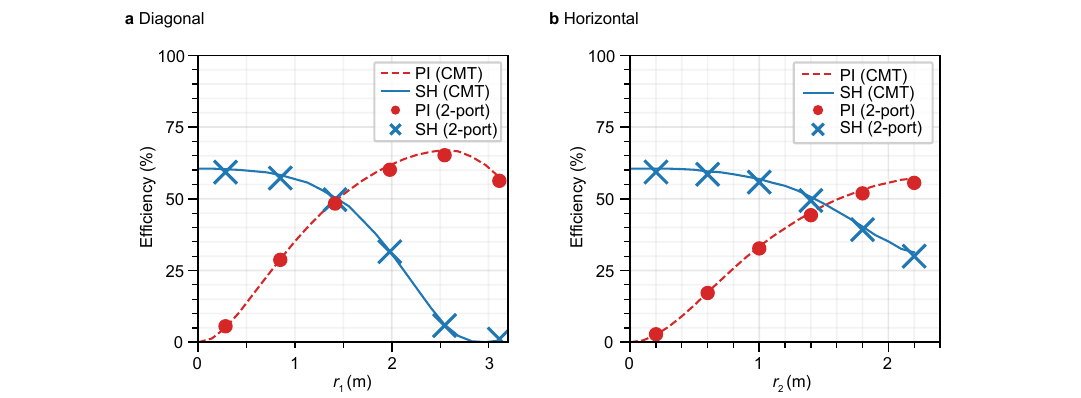}
\caption{Validation of coupled-mode theory (CMT) against two-port full-wave simulations along representative receiver trajectories shown in Fig.~\ref{fig_excitation}a. (a) Power-transfer efficiency along the diagonal path. (b) Power-transfer efficiency along the horizontal path. The agreement between the CMT predictions and the two-port simulation results for the PI and SH modes confirms the accuracy of the proposed volumetric-efficiency model.}
\label{fig_validation}
\end{figure}

In the S-parameter-based validation, the RX is translated along the representative diagonal and horizontal trajectories shown in Fig.~\ref{fig_excitation}a (see Methods). The resulting efficiency profiles are presented in Figs.~\ref{fig_validation}a and \ref{fig_validation}b alongside the CMT predictions. Along the diagonal path, the SH mode is dominant near the central region, whereas the PI mode becomes dominant toward the periphery. A similar crossover is observed along the horizontal path. In both cases, the CMT predictions closely agree with the full-wave simulation results, validating the accuracy of the CMT-based volumetric efficiency analysis.

\section*{Discussion}
In this work, we presented a patched-wall quasistatic cavity resonator for room-scale wireless power transfer that eliminates the need for obstructive internal conductive structures while preserving full-volume coverage. The proposed architecture supports two complementary resonant modes: a pole-independent (PI) mode that maintains coverage near the room boundaries, and a surface-Helmholtz (SH) mode that generates strong magnetic fields in the central region without requiring a conductive pole. Volumetric efficiency analysis, supported by full-wave validation along representative receiver trajectories, shows that the proposed pole-free architecture provides wide-area three-dimensional wireless power coverage while preserving an unobstructed interior space. As summarized in Table~\ref{tab:simple_comparison}, the proposed patched-wall QSCR retains the full-volume coverage capability of prior dual-mode multimode QSCR while removing the need for an internal conductive structure. Although this study focuses on simulation-based evaluation, the agreement between the CMT predictions and full-wave results supports the proposed design.

\begin{table}[t]
\centering
\begin{threeparttable}
\caption{Qualitative comparison of room-scale WPT resonators.}
\label{tab:simple_comparison}
\renewcommand{\arraystretch}{1.25}
\begin{tabular}{lcccc}
\toprule
 & \multicolumn{3}{c}{Prior work} & This work \\
\cmidrule(lr){2-4} \cmidrule(l){5-5}
 & QSCR\tnote{a} & M-QSCR\tnote{b} & M-QSCR\tnote{b} & PW-QSCR\tnote{e} \\
 &  & (PI\tnote{c} ) & (Dual\tnote{d} ) & (Dual\tnote{d} ) \\
\midrule
Full-volume & $\times$ & $\times$ & $\checkmark$ & $\checkmark$ \\
Unobstructed & $\times$ & $\checkmark$ & $\times$     & $\checkmark$ \\
\bottomrule
\end{tabular}
\begin{tablenotes}[flushleft]
\footnotesize
\item[a] QSCR: quasistatic cavity resonator~\cite{Chabalko2017}.
\item[b] M-QSCR: multimode quasistatic cavity resonator~\cite{sasatani_multimode_2017, sasataniRoomscaleMagnetoquasistaticWireless2021}.
\item[c] PI: pole-independent mode. PI mode-only operation in prior M-QSCR can be realized without the central pole, but full-volume coverage in prior work requires dual-mode operation.
\item[d] Dual: Dual-mode operation.
\item[e] PW-QSCR: patched-wall quasistatic cavity resonator.
\end{tablenotes}
\end{threeparttable}
\end{table}

The proposed architecture builds on components and construction practices already demonstrated in prior QSCR systems. Room-scale QSCR prototypes have previously been constructed from conductive wall panels and lumped capacitors~\cite{Chabalko2017,sasataniRoomscaleMagnetoquasistaticWireless2021}, and the patched-wall structure relies on the same class of components: planar conductive sheets and capacitors whose required capacitance and quality factor are consistent with commercially available multilayer ceramic capacitors~\cite{kyocera_avx_600s}. For deployment, safety analyses of prior QSCR implementations indicate that useful power levels can be delivered while complying with established exposure guidelines~\cite{Chabalko2017, icnirp_guidelines_2020}, and a corresponding exposure characterization of the patched-wall structure is an important next step.

Several limitations of the present study should be noted. Our contribution is the resonator structure and its complementary mode pair, evaluated through finite-element simulation; experimental demonstration is left to future work. The simulation methodology employed here, however, builds on modeling approaches that have been validated against experiments in prior cavity-based WPT systems, including the multimode QSCR demonstrated at room scale~\cite{sasatani_multimode_2017,sasataniRoomscaleMagnetoquasistaticWireless2021,sasatani_geometry-based_2022}, and the close agreement between the CMT predictions and independent full-wave two-port simulations supports the internal consistency of the reported efficiency analysis. In addition, the volumetric analysis assumes an ideally matched load at each receiver position, which is a common practice for characterizing wireless power resonators~\cite{Chabalko2017,sasataniRoomscaleMagnetoquasistaticWireless2021}, as well as an optimally oriented receiver. The simulation data and analysis code that reproduce the reported results are openly available (see Data availability).

Eliminating internal conductive structures removes a key barrier to deploying room-scale WPT in everyday environments. Unobstructed full-volume coverage would allow mobile and IoT devices to be powered anywhere within the volume without alignment to charging pads or surfaces, and could support emerging applications such as untethered sensors~\cite{akyildiz_wireless_2002, xu_internet_2014} and actuators~\cite{Boyvat2017, Lee2011}, as well as experimental systems in instrumented enclosures~\cite{Montgomery2015, aharoni2026wireless}. More broadly, the surface-Helmholtz mode illustrates that structured boundary currents can synthesize interior field distributions that previously required internal conductors, suggesting a wider design space for cavity-based WPT resonators.

\section*{Methods}

\subsection*{Finite-element eigenmode simulations}
We used the RF module in COMSOL Multiphysics to model the enclosure with 1 mm-thick A1100 aluminum sheets and performed eigenvalue simulations to extract the resonant frequencies, quality factors, and field distributions of the two supported modes. The nominal capacitance was set to $C=11$ nF with an assumed quality factor of 2000, consistent with recent specifications for high-$Q$ multilayer ceramic capacitors~\cite{kyocera_avx_600s}.

\subsection*{Volumetric efficiency analysis based on coupled-mode theory}
Coupled-mode theory (CMT) is employed as an approximate model in which the coupling coefficient is calculated from the total magnetic energy and the magnetic flux intersecting the RX resonator, while the modal quality factors are extracted from the COMSOL eigenvalue simulations. The RX resonator is modeled as a 6-turn, $165 \times 165$~mm$^2$ square helix with a quality factor ($Q$) of 340 and an inductance of $13.6$~$\mu$H. For both the CMT analysis and the S-parameter-based validation, the load impedance is assumed to be optimized to maximize the power transfer efficiency at each receiver position~\cite{Chabalko2017,sasatani_multimode_2017}.

In this framework, the maximum efficiency $\eta_{\rm max}$ achievable between a pair of resonators with quality factors $Q_1$ and $Q_2$ and coupling coefficient $k_{1,2}$ is determined by the dimensionless parameter $k_{1,2}\sqrt{Q_1Q_2}$ (the $kQ$-product)~\cite{Zargham2012, Haus1991}:

\begin{equation}
\eta_{\rm max} = \dfrac{k_{1,2}^2 Q_1 Q_2}{\left(1+\sqrt{1+k_{1,2}^2 Q_1 Q_2}\right)^2}
\label{equation:max_eff}
\end{equation}

The coupling coefficient $k_{1,2}$ is physically related to the magnetic flux linkage $\beta$ through the receiver and the total magnetic energy stored in the cavity volume $\alpha$~\cite{chabalko_resonant_2014}:

\begin{equation}
k_{1,2} = \dfrac{\beta}{\sqrt{2L_2\alpha}}
\label{eq:kappa}
\end{equation}
\begin{equation}
\beta = \iint_{A_2} \left(\mu_0\vec{H}_1\cdot\vec{n}\right)\,\mathrm{d}A_2
\label{eq:beta}
\end{equation}
\begin{equation}
\alpha = \iiint_{V_1} \left(\mu_0\dfrac{|\vec{H}_1|^2}{2}\right)\,\mathrm{d}V_1
\label{eq:alpha}
\end{equation}
Here, $\vec{H}_1$ denotes the magnetic field of the cavity mode, $\vec{n}$ is the unit normal to the RX coil plane, $A_2$ and $L_2$ are the area and inductance of the RX coil, and $V_1$ is the cavity interior volume.

For the volumetric analysis, the RX is sampled across 0.1~m grid points within the interior volume, defined by $|x|, |y| \leq 2.2$~m and $|z| \leq 1.0$~m. Because the room-scale resonator is approximately 500 times larger than the RX, the magnetic field is assumed to be uniform across the RX coil cross-section. The RX orientation is then assumed to be locally aligned to maximize flux linkage, such that the coupling is approximated from the local magnetic-field magnitude and the RX coil area. The dual-mode efficiency is defined as the larger of the SH- and PI-mode efficiencies at each receiver position, corresponding to a system that switches between the two modes according to the receiver location.

\subsection*{Full-wave validation}
To verify the accuracy of the CMT analysis, full-wave two-port S-parameter simulations are performed for a subset of discrete receiver positions. The source is stimulated at ports $P_{\mathrm{SH}}$ and $P_{\mathrm{PI}}$ (Fig.~\ref{fig_excitation}a) to excite the SH and PI modes, respectively, and the efficiency is evaluated with the load impedance optimized at each receiver position~\cite{Zargham2012}.

\subsection*{Use of generative AI}
Generative AI tools (OpenAI GPT, Google Gemini, and Anthropic Claude) were used to assist in the development of the analysis software and in the drafting and refinement of the manuscript.

\bibliography{sample}

\section*{Acknowledgements}
This work was supported by JST FOREST (Grant Number JPMJFR242P), JST ACT-X (Grant Number JPMJAX190F), and JSPS KAKENHI (Grant Number 23K28068).

\section*{Author contributions}

T.S. designed the research, conceived the simulation, analyzed the results, and wrote the manuscript. Y.K. reviewed and commented on the manuscript. All authors reviewed and approved the final manuscript.

\section*{Additional information}

\subsection*{Data availability}
The simulation data and analysis code that support the findings of this study are openly available in the Zenodo repository at \url{https://doi.org/10.5281/zenodo.19257915}~\cite{sasatani_2026_19257915}, and are also available on GitHub at \url{https://github.com/SasataniLab/PW-QSCR}.

\subsection*{Competing interests}
The authors declare no competing interests.

\end{document}